\documentclass[reprint,amsmath,amssymb,notitlepage]{revtex4-1}
\pdfoutput=1

\usepackage{graphicx}
\usepackage{hyperref}

\makeatletter
\hypersetup{
	    pdfcreator={Cesar Henrique Comin},
		colorlinks=true,       		
    	linkcolor=black,          	
    	citecolor=black,        		
    	filecolor=black,      		
		urlcolor=black,
		bookmarksdepth=4
}
\makeatother

\begin{document}

\title{A framework for evaluating complex networks measurements}

\author{Cesar H. Comin$^1$}
\email{Email: chcomin@gmail.com} 
\author{Filipi N. Silva$^1$}
\author{Luciano da F. Costa$^1$}

\affiliation{$^1$S\~ao Carlos Institute of Physics, University of S\~ao Paulo, S\~ao Carlos, S\~ao Paulo, Brazil}

\begin{abstract}

A good deal of current research in complex networks involves the characterization and/or classification of the topological properties of given structures, which has motivated several respective measurements. This letter proposes a framework for evaluating the quality of complex network measurements in terms of their \emph{effective resolution}, \emph{degree of degeneracy} and \emph{discriminability}. The potential of the suggested approach is illustrated with respect to comparing the characterization of several model and real-world networks by using concentric and symmetry measurements. The results indicate a markedly superior performance for the latter type of mapping.

\end{abstract}

\maketitle

\section{Introduction}
\begin{figure}[]
    \includegraphics[width=\linewidth]{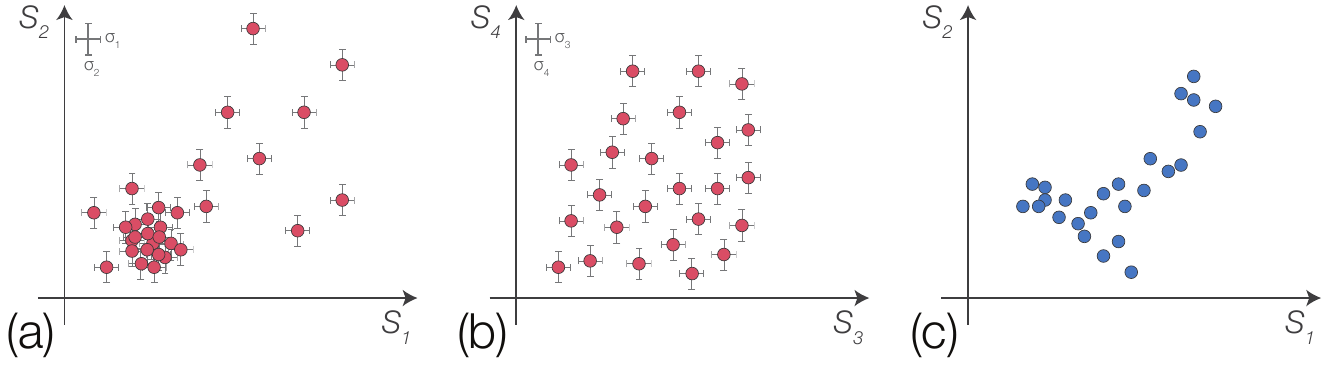}
    \caption{Illustration of the distinct distributions that can be observed depending on the choice of measurements or entities under analysis. (a) Heterogeneous distribution. (b) A more uniform distribution for the same entities as in (a) obtained by using different measurements. (c) A distinct distribution observed for the same measurements as in (a) but a distinct set of entities. The errorbars represent the level of noise of each respective measurement.}
    \label{f:motiv}
\end{figure}

Science is based on the objective quantification of properties of the phenomenon under analysis. Though it is frequently impossible to use a complete set of measurements, so as to allow the phenomenon to be reconstructed, it is expected that a good set of measurements would be able to provide a comprehensive characterization of the relevant properties. 
Typically, the characterization of a phenomenon involving several \emph{entities}, such as objects or instances of an object (e.g., along time), requires the selection of one or more measurements of them. Once these measurements have been chosen, the obtained values of the measurements define a distribution of points in the respective measurement space. For instance, suppose that a given set of entities is characterized by measurements $S_1$ and $S_2$. A possible distribution of those measurements is shown in fig.~\ref{f:motiv}(a). If we were to apply two other measurements to the same set of entities, the resulting distribution could be markedly distinct, as shown in fig.~\ref{f:motiv}(b). That is, the same set of entities can produce completely different distribution of points for different choices of measurements. At the same time, two distinct sets of entities typically yield different distributions of points for the same measurements, as illustrated in fig.~\ref{f:motiv}(c). Therefore, the distribution of points in the measurements space is a consequence of both the specific set of entities under analysis and the choice of measurements.

The potential of a measurement to represent a set of entities can be reduced to three mains aspects, namely the uniform resolution of the measurement, the degree of degeneracy of the mapping and the intrinsic discriminability of the measurement to different categories of data, i.e.\ its performance in classification. Regarding \emph{resolution}, it can be related to the degree of accuracy (e.g. numerical error) that can be achieved and to how well distributed the entities result in the measurement space, so that fixed resolution of measurement is achieved. For instance, the presence of clusters of entities with similar values intrinsically implies in voids between such clusters. By \emph{degeneracy} of a measurement it is often understood the loss of information while mapping the entity into a set of values, so that it cannot be recovered from this set. In other words, a degenerate mapping is non-invertible. A simple example of a degenerate measurement is the degree of a node in a graph, i.e.\ the original graph cannot be recovered from its degree distribution. The \emph{discriminability} of a measurement is important for the classification of the entities in the sense that entities of the same type result in similar measurement values. 

In order to quantify the aforementioned aspects, we define the \emph{measurement evenness} and \emph{exclusion} properties. A measurement that has high evenness value is more uniformly distributed over the measurement regions. In cases when the measurement needs to be binned, the enhanced uniformity will promote a more effective use of the bins. Contrariwise, a less uniformly distributed measurement would require adaptive binning such that smaller bins are allocated to the higher density regions, which is difficult to achieve in practice. In addition, a more uniform measurement will be more robust to noise and perturbations affecting the mapping of the entities in the measurement space. For instance, several real-world data are incompletely sampled. This is achieved because a more uniform distribution of points will tend to occupy the space more effectively, avoiding gaps and therefore providing a larger average distance between adjacent pairs of points. This is illustrated in fig.~\ref{f:motiv}. If the magnitude of noise $\sigma_1$ and $\sigma_2$ at each axis is as shown in fig.~\ref{f:motiv}, the results in fig.~\ref{f:motiv}(a) will be completely undermined at the higher density region, while the distribution of points in fig.~\ref{f:motiv}(b) would be much less affected.

A measurement that can correctly represent the entities of the system must also be sensitive to differences in the types of entities under analyses. For this task, we define the exclusion property, which quantifies the mixing of different classes of data in the measurement space. Therefore, in this work, the requirement of having high evenness, implying in having a more uniform distribution of points, and high exclusion allows for better binning, robustness to perturbation and noise and better discriminability of the data.

We will develop the aforementioned ideas using complex networks as the set of entities under analysis. The area of Complex Networks~\cite{Barabasi99:Science,newman2003thestructure} has grown steadily since its origin in 1999, mainly as a consequence of its ability to represent virtually any discrete system~\cite{Costa11:AP}. Basically, these networks are graphs exhibiting a topological organization which departs from a randomly uniform network such as Erd\H{o}s-R\'enyi~\cite{newman2003thestructure}, which acts as a ``simple" reference. The study of complex networks involves the estimation of several measurements, such as the degree, clustering coefficient, and betweenness centrality. Basically, network measurements can be classified as being \emph{global}, such as the node degree distribution, or \emph{local}, referring to small parts of the network. Thus, the topological properties around each node are mapped into a set of values, allowing a comprehensive approach to determining the aspects that are similar or different between two or more networks. However, because of the three problems identified in the previous paragraph, this approach requires good quality measurements. This corresponds to the objective of the present work, i.e.\ we propose a framework for assessing the quality of different sets of node-centered measurements of complex networks with respect to \emph{effective resolution}, \emph{degree of degeneracy}, and \emph{discriminability}.

\section{Methodology}

The first step in our methodology is to define a set of node-centered measurements that will be used to characterize the networks. Such measurements are calculated over connectivity patterns along the neighborhood of nodes~\cite{Costa:2006p278}. Where a $l$-th neighborhood is defined as the set of nodes that are at a topological distance $l$ from a reference node. The subgraph spanned by the first $r$ neighborhoods of a node is henceforth called the $r$-pattern of the node. We note that, depending on the network type and on its average degree, the $r$-pattern of nodes can span the entire network even for small values of $r$. Nevertheless, such patterns will still be usually distinct one another since they have a hierarchical structure, composed by the successive neighborhoods of the reference node.

\begin{figure}[]
    \includegraphics[width=\linewidth]{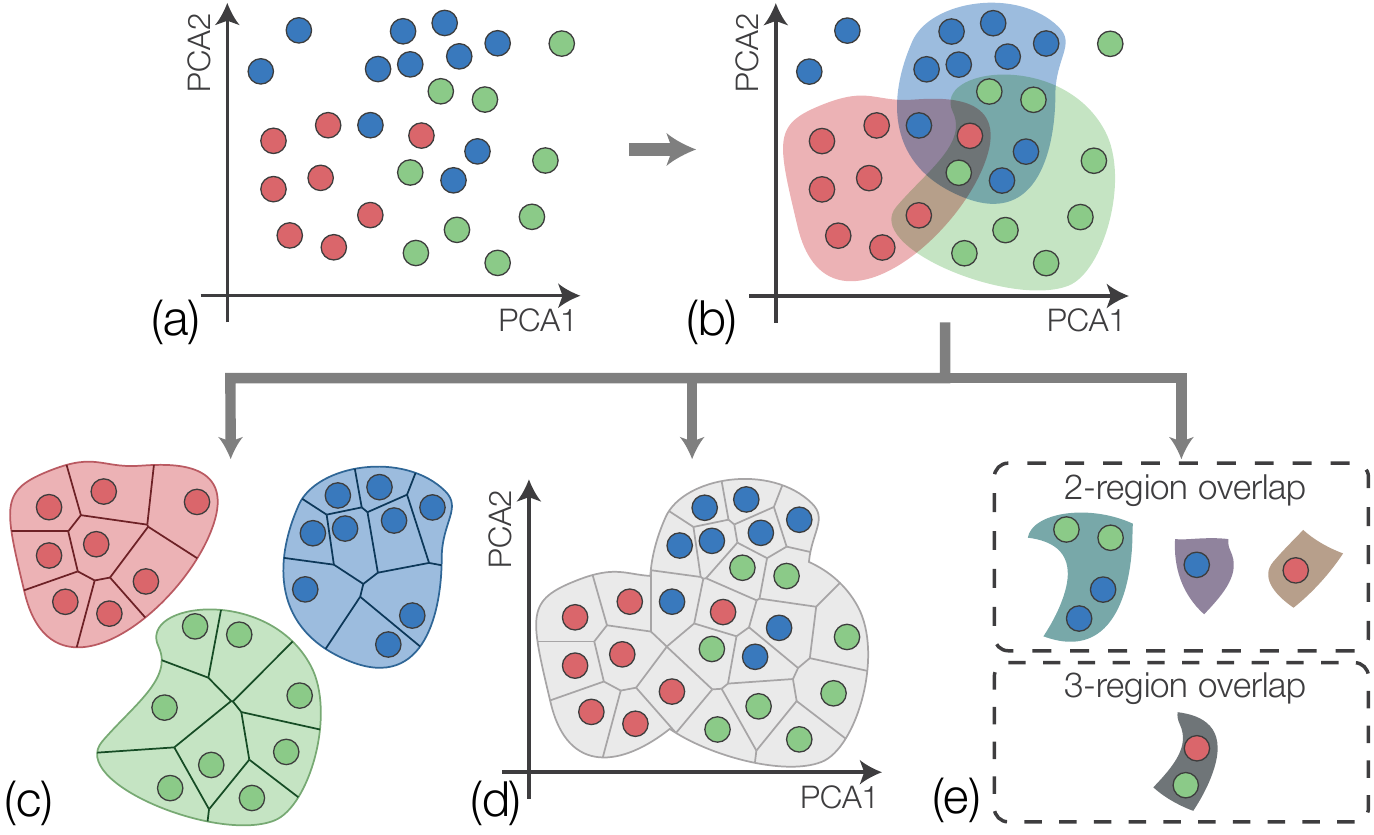}
    \caption{Steps of the proposed methodology.}
    \label{f:methodology}
\end{figure}

The following steps can be applied to the original hyperdimensional space composed by a large number of node-centered measurements. Nevertheless, in order to provide a visual interpretation of the methodology, and to reduce its computational cost, we apply Principal Component Analysis (PCA)~\cite{jolliffe:2002} on the data. Using PCA, we can project the original measurements into a 2D space, composed by the first two principal components. In fig.~\ref{f:methodology}(a) we show an example of such 2D space, where patterns are represented by points projected on the first two principal components obtained from a set of node centered measurements. Patterns are colored according to the network they belong to.

If the $r$-patterns of two given nodes are the same, no measurement will be able to differentiate the two nodes. Therefore, in order to quantify the potential of a set of measurements to characterize networks, we need to take into account identical patterns, so that they will be considered only once. This is done by finding isomorphic patterns between all nodes contained in the projection, regardless of the network they belong to. The isomorphic patterns are found by first coloring the vertices of the patterns according to their distance from the reference vertex, and then considering the color-preserving isomorphism~\cite{scapellato2003topics} between all patterns. 

The data presented in fig.~\ref{f:methodology}(a) is used to define the typical measurement region of each network. This is done by applying a kernel density estimation~\cite{Silverman1998density} to the projected data of each network, where a normal distribution is used as the kernel. Then, a threshold, $T_c$, is applied to the estimated density for each network $c$. Regions of the space having density values higher than $T_c$ are defined as measurement regions of the network. The value of $T_c$ is set so that the sum of the estimated probability density inside the measurement region is as close as possible to a given fixed value $p$. This is done to eliminate outliers in the data, as well as have regions containing a similar number of nodes. The procedure has two free parameters, which are the bandwidth, or standard deviation, of the normal distribution used as the kernel and the value $p$. These parameters are not critical to the method, provided that the same values are used for both sets of measurements that are being compared. We note that if a point is related to an isomorphic pattern, in the kernel density estimation we consider this point as many times as the number of isomorphic patterns it represents. In fig.~\ref{f:methodology}(b) we show the resulting regions defined by the patterns in fig.~\ref{f:methodology}(a). A good discriminative measurement should have two main characteristics. First, the non-isomorphic patterns belonging to a single network must be as disperse as possible in the measurement space. That is, distinct patterns must show distinct measurement values. Second, the overlap between distinct network regions must be as small as possible, reducing the probability of classification errors. A first approach to quantify the dispersion of the points would be to measure the areas of the regions. However, these areas do not posses information about the distribution of the points inside each region. We define a more powerful measurement of the point spread of a region, which we call \emph{evenness}.

The evenness is closely related to the accessibility of networks, which provides the effective degree of nodes~\cite{viana2012effective}. The measurement is calculated as follows. First, we find the Voronoi tessellation~\cite{Ahuja1983pattern} of each region separately, where the points define the position of the Voronoi cells. An example is shown in fig.~\ref{f:methodology}(c). Second, the Voronoi tessellation is used to define a fractional area distribution $P(A_i)$, which is given by the area of the Voronoi cells divided by the area of the entire region. Note that this is different from taking the area distribution of the cells. Since we are working with unique patterns, when two points have exactly the same position in the space it means that the patterns were distinct, but the measurements were not able to distinguish them. We consider that all points having the same position define a single Voronoi cell. Then, the entropy of the fractional area distribution of the region is calculated as
\begin{equation}
E_c=-\sum\limits_{i=1}^{N_p} P(A_i)\log (P(A_i)),
\end{equation}
where $N_p$ is the number of unique patterns inside the region. If the measurement set were able to distinguish all non-isomorphic patterns, and the cell areas were all equal, then $E_c=\log (N_p)$. Therefore, we define the region evenness as 
\begin{equation}
\zeta_c=\frac{e^{E_c}}{N_p},
\end{equation}
which has values in the range $[1/N_p,1]$. An overall region evenness can be found by taking the average of the region evenness for all networks. Another way to define an overall region evenness is by constructing a Voronoi tessellation of all regions together, without distinguishing between regions. An example of such Voronoi tessellation is shown in fig.~\ref{f:methodology}(d).

\begin{figure*}[!t]
    \includegraphics[width=0.9\linewidth]{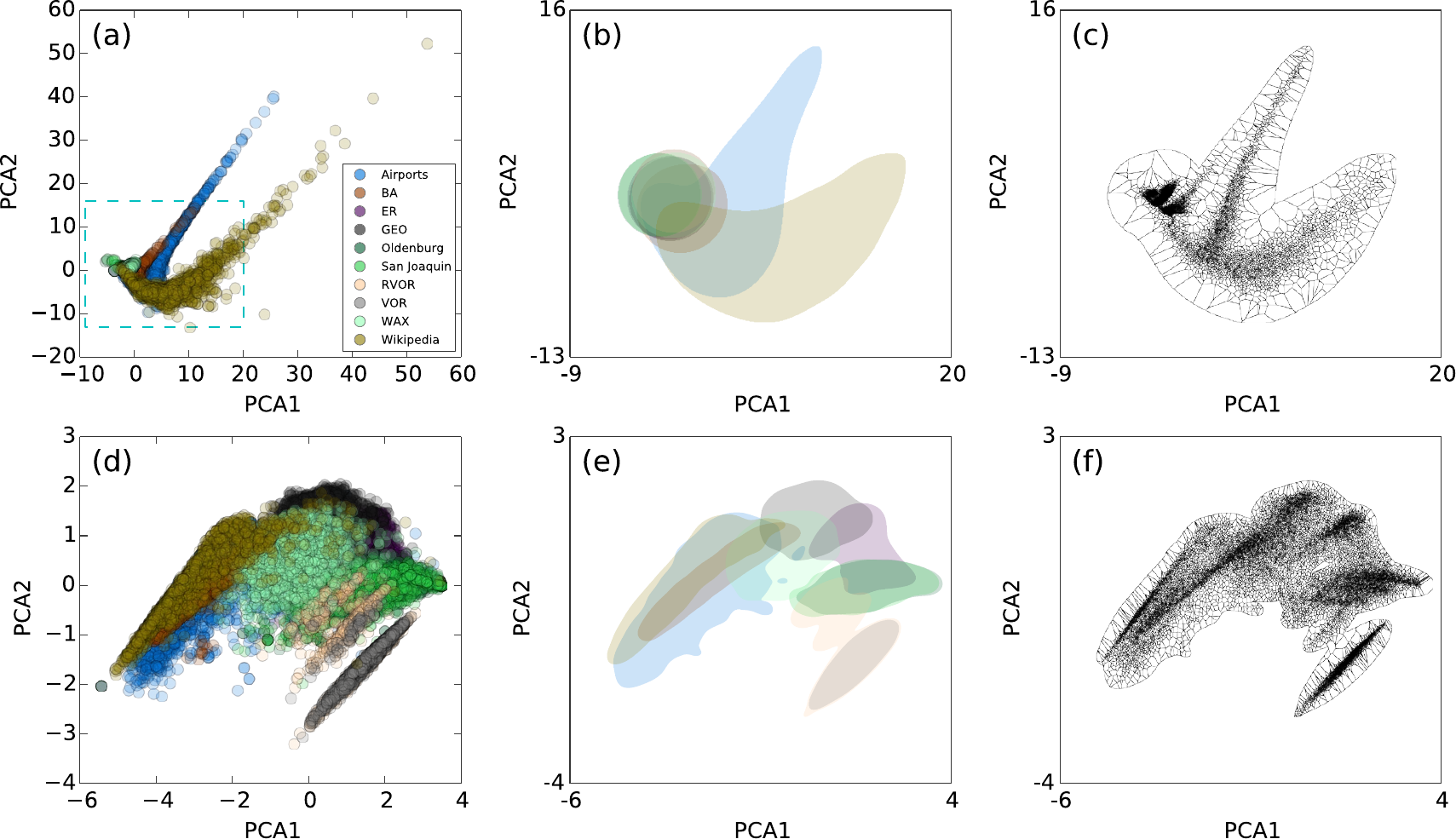}
    \caption{Results of each step of the methodology for the concentric (first row) and symmetry (second row) measurement sets.}
    \label{f:results}
\end{figure*}

As stated above, the second property of a representative set of measurements is that the overlaps between regions must be minimal. We quantify the overlaps by counting the number of excess regions that each pattern belongs. An excess region of a pattern is defined as the number of regions that a pattern falls into, minus the number of distinct networks that the pattern belongs. For example, suppose that a given pattern can be found in three distinct networks, but the point respective to that pattern falls into four distinct regions. Then, the number of excess regions of the pattern is one. The sum of excess regions of all points define what we call the overlap, $V$, between regions. In fig.~\ref{f:methodology}(e) we show the overlapping regions defined in fig.~\ref{f:methodology}(b), together with the respective unique patterns contained in these regions. The upper bound, $V_{\mathrm{max}}$, of the overlap between regions is attained when all unique patterns fall into all regions. Therefore, in order to quantify the discriminability of a set of measurements we define the \emph{exclusion} of the set, which is given by

\begin{equation}
\xi=1-V/V_{\mathrm{max}}.
\end{equation}

\section{A case example}

Our methodology was applied to 10 networks having markedly distinct characteristics, six of which are generated from network models and four are real-world networks. In table~\ref{tab:nets} we show some basic characteristics of the networks, as well as key references about them. The models used to generate the networks were the Erd\H{o}s-R\'enyi (ER), Barab\'asi-Albert (BA), Random Geometric (GEO), Waxman (WAX), Voronoi (VOR) and rewired Voronoi (RVOR). The first four models are well-known in the complex network literature and their precise definition can be found in the supplied references. The Voronoi model is defined through the Voronoi tessellation~\cite{Ahuja1983pattern} of a set of points. We start by placing randomly, with uniform probability, a set of $N$ points in a 2D space. Then, a Voronoi tessellation of the points is created and each node is associated with a Voronoi cell. Nodes having adjacent Voronoi cells are connected, thus defining a Voronoi network. The rewired Voronoi model is defined by applying a random rewiring of a Voronoi network, were the probability of rewiring is 0.001. It is important to note that the six models differ mainly by the spatial constraints imposed on the network creation. While the ER and BA models have no spatial constraints, the WAX, GEO, RVOR and VOR models have progressively stricter constraints in the allowed number of crossing between network edges. The four real-world networks are the World-Wide Airport network (Airport), the Wikipedia and the street networks of the city of Oldenburg (Oldenburg) and the county of San Joaquin (San Joaquin). 

\begin{table}[]
\caption{\label{tab:nets}Number of nodes, $N$, and average degree, $\langle k\rangle$, of the networks used in the main paper. Key references describing the networks are indicated in the last column.}
\begin{center}
\begin{tabular}{cccc}
\hline
\textbf{Network} & $N$ & $\langle k\rangle$ & Ref. \\
\hline
\textbf{Airports} & 2940 & 20.9 & \cite{Silva2014concentric} \\
\textbf{BA} & 5000 & 6.00 & \cite{Barabasi99:Science} \\
\textbf{ER} & 5000 & 6.07 & \cite{newman2003thestructure} \\
\textbf{GEO} & 4964 & 5.77 & \cite{barthelemy2011spatial} \\
\textbf{Oldenburg} & 2873 & 2.64 & \cite{Brinkhoff2002} \\
\textbf{San Joaquin} & 14503 & 2.77 & \cite{Brinkhoff2002} \\
\textbf{RVOR} & 5000 & 5.99 & \cite{Silva2014concentric} \\
\textbf{VOR} & 5000 & 5.99 & \cite{costa2004voronoi,barthelemy2011spatial} \\
\textbf{WAX} & 5000 & 6.04& \cite{waxman1988routing,barthelemy2011spatial} \\
\textbf{Wikipedia} & 45876 & 11.77 & \cite{Silva2014concentric} \\
\hline
\end{tabular}
\end{center}
\end{table}

Since the networks have markedly distinct number of nodes, we randomly selected $N_s=2000$ nodes from each network, so as that they all have the same relevance in the PCA. We verified that applying the PCA to different sets of randomly selected nodes represented unnoticeable changes to the results. Two sets of measurements were used to characterize the neighborhoods of nodes, namely concentric measurements~\cite{Costa:2006p278,Costa:2008p14} and symmetry measurements~\cite{Silva2014concentric}. Concentric measurements are simple statistics of the neighborhood of nodes, such as the number of nodes at the $i$-th neighborhood or the number of edges between successive neighborhoods of a node.  They are related to many traditional measurements in network theory, as described in~\cite{Silva2014concentric}. Symmetry measurements quantify the topological symmetry of the nodes neighborhoods. They correspond to a normalization of the accessibility measurement~\cite{viana2012effective} and have been found to provide a rich description of the topological structure of networks \cite{Silva2014concentric}. 

Starting with the concentric measurements, in fig.~\ref{f:results}(a) we show the PCA projection of all concentric measurements presented in \cite{Costa:2006p278}, for the 0, 1, 2, 3 and 4-th neighborhoods. It is clear that most of the network models became concentrated in a small region, around the origin of the axes. Only the Airport and Wiki networks contain nodes having more distinct values of PCA1 and PCA2, although the 2D space is still poorly occupied by the two networks. The main reason for this behavior is that the concentric measurements present different scales depending on the network characteristics. For the Airport and Wiki, which are highly heterogeneous networks, these measurements show markedly distinct values, which is an indication of a good measurement according to our criteria. But all other networks are poorly characterized by the concentric properties. Therefore, we expect to obtain low values of evenness and exclusion from our methodology. The regions defined by the PCA projection are shown in fig.~\ref{f:results}(b) for $p=0.9$. Note that we show a zoomed in version of the PCA axes, indicated by the blue dashed line in fig.~\ref{f:results}(a). We notice that the regions defined for eight of the networks are strongly overlapping. The respective Voronoi tessellation considering all regions is shown in fig.~\ref{f:results}(c), where the highly heterogeneous distribution of cell areas is evident. The global evenness and exclusion measurements obtained from the concentric measurements are shown in table~\ref{tab:results}. In the table, we also show the standard deviation of the measurements over 10 realizations of the methodology when sampling distinct sets of nodes from each network.

\begin{figure}[!tb]
    \includegraphics[width=\linewidth]{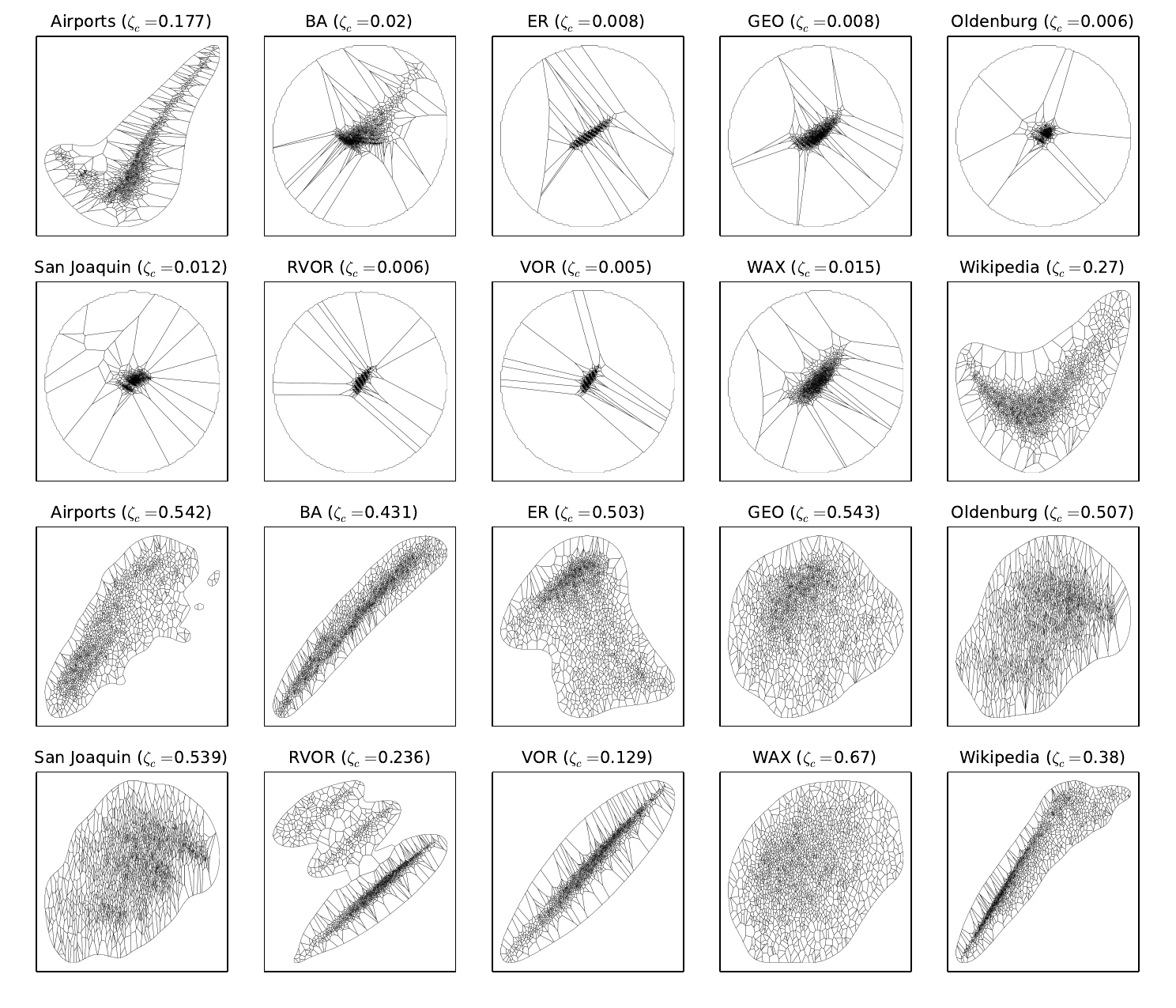}
    \caption{Voronoi tessellation of each network region respective to concentric (first and second rows) and symmetry (third and fourth rows) measurements. The evenness for each region is shown above the respective plot.}
    \label{f:voronoi}
\end{figure}

We compare the results obtained for the concentric measurements with those attained by the symmetry measurements. The symmetry measurements described in \cite{Silva2014concentric} were applied to the 2, 3 and 4-th neighborhoods of each node of our set of networks. The resulting PCA projection is shown in fig.~\ref{f:results}(d). It is immediately clear that the symmetry measurements provided a more uniform filling of the 2D PCA projection. All networks seem to be characterized on the same scale, and they all appear to belong to a well-defined region of the measurement space. This constitutes a good set of measurements, according to our criteria. The respective regions defined by the kernel density estimation are shown in fig.~\ref{f:results}(e). We note that there are many non-overlapping network regions. The global Voronoi tessellation is shown in fig.~\ref{f:results}(f). Throughout most of the defined region, cell sizes are highly homogeneous, with the exception of border regions in the Voronoi models and the Wikipedia. These border regions present such variation of cell sizes because of the high concentration of points in a small region for these networks. In table~\ref{tab:results} we show the global evenness and exclusion obtained for the symmetry measurements set.

As explained above, another way to quantify the evenness of the measurement set is by taking the evenness of each region separately. In fig.~\ref{f:voronoi} we show the Voronoi tessellation for each network region separately. The first two lines of images are related to the concentric set of measurements. It is clear that, again, only the Airport and Wikipedia networks were well described by the measurements. The other networks present highly distinct cell areas. As for the symmetry, most regions show an homogeneous distribution of cell areas, with the exception of the VOR model and, to some extent, the RVOR and Wikipedia networks. The  average evenness calculated separately for each region of each set of measurements is indicated in table~\ref{tab:results}. 

\begin{table}[!h]
\caption{\label{tab:results} Exclusion and evenness values obtained for the concentric and symmetry measurements studied in this work. Each measurement is calculated over 10 realizations of the methodology, where each realization is based on a different set of sampled nodes from the networks.}
\begin{center}
\begin{tabular}{ccc}
\hline
 & \textbf{Concentric} & \textbf{Symmetry} \\
\hline 
\textbf{Exclusion} & $0.237 \pm 0.047 $ & $0.848 \pm 0.002$  \\
\textbf{Global evenness} & $0.048 \pm 0.014$ & $0.387 \pm 0.005$ \\
\textbf{Average evenness} & $0.048 \pm 0.013$ & $0.447 \pm 0.003$ \\
\hline
\end{tabular}
\end{center}
\end{table} 

\begin{figure}[]
    \includegraphics[width=\linewidth]{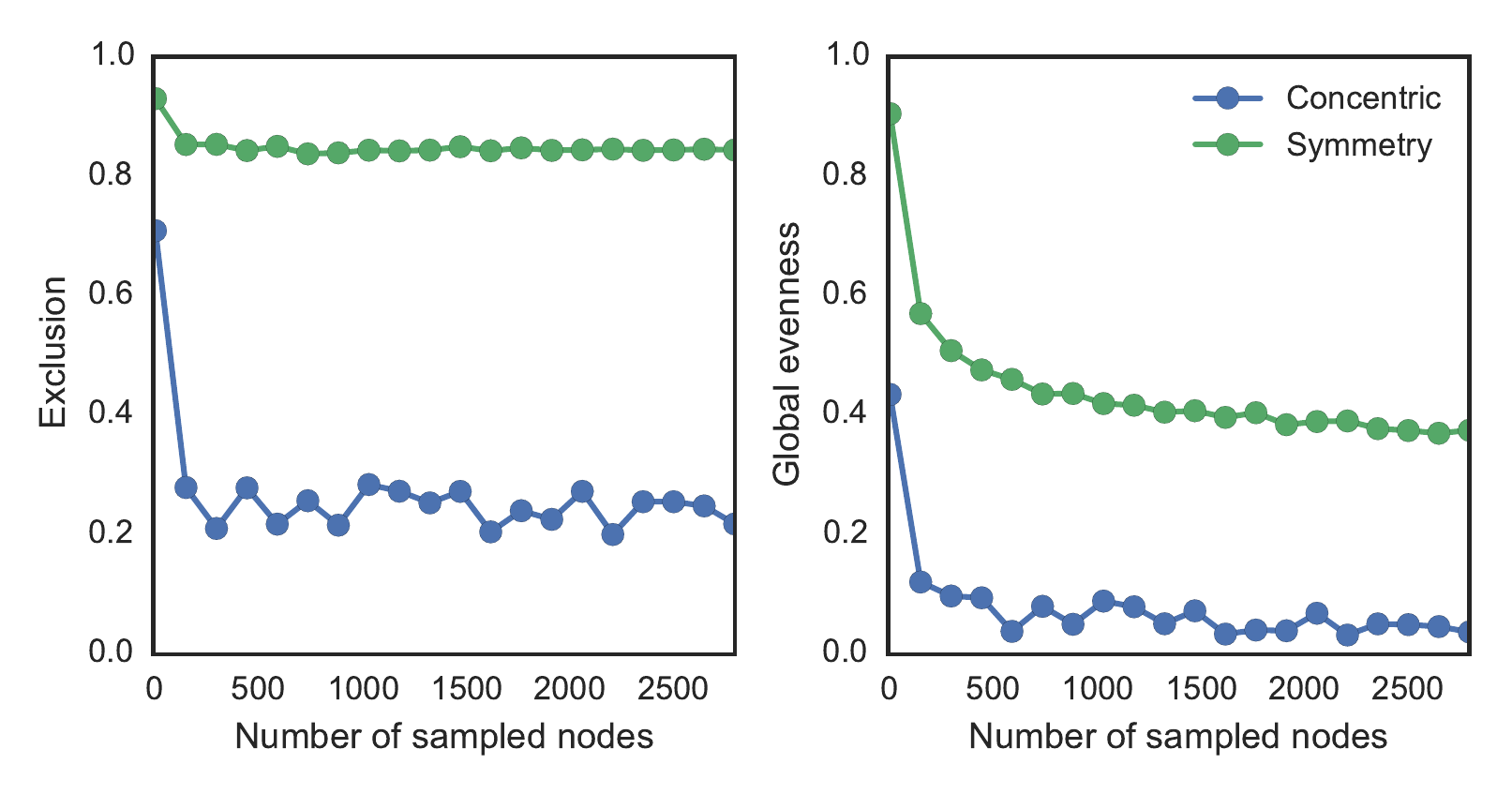}
    \caption{Exclusion and evenness of the concentric and symmetry measurements as a function of the number of sampled nodes from the networks.}
    \label{f:varynumnodes}
\end{figure}

An important step of the methodology is the sampling of the patterns used to characterize the networks. We presented the results for a sampling of $N_s=2000$ patterns, but it is important to verify if changing $N_s$ can influence the results. In fig.~\ref{f:varynumnodes} we show the exclusion and evenness measurements as a function of the number of sampled nodes from each network. It is clear that the relative quality between the concentric and symmetry measurements sets does not significantly change when more than 200 nodes are sampled. Therefore, the measurement sets can be correctly evaluated using only a small number of patterns from each network.  This represents sampling 7\% of the smallest network (Oldenburg streets) and 0.4\% of the largest one (Wikipedia).

\begin{figure}[]
    \includegraphics[width=\linewidth]{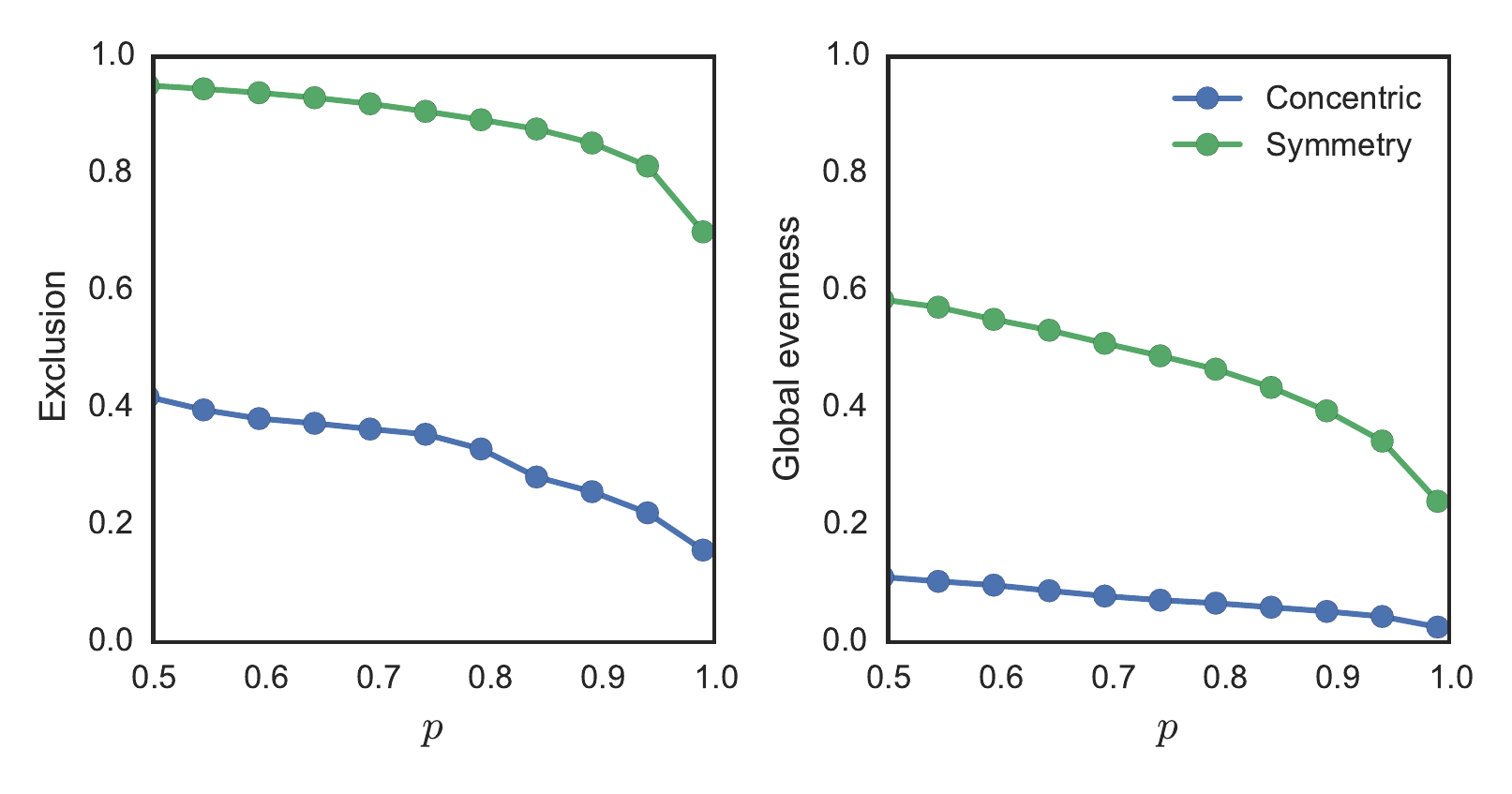}
    \caption{Exclusion and evenness of the concentric and symmetry measurements as a function of the parameter $p$, used for thresholding the probability density of the measurements.}
    \label{f:varyp}
\end{figure}

The parameter $p$ used to define the threshold of the probability density for each network could also influence the results. If this parameter is too high, the measurement regions of the networks, shown in fig.~\ref{f:results}(b) and (e), become strongly influenced by outliers in the PCA projection. In order to demonstrate the influence of $p$, in fig.~\ref{f:varyp} we show the exclusion and evenness of the concentric and symmetry measurements as a function of $p$. The relative value of the exclusion for the two measurement sets has little change even for values of $p$ approaching 1. The evenness presents more significant changes, especially for $p>0.95$, but the evenness difference between the concentric and symmetry measurements remains large for most values of $p$.

\section{Conclusion}

The results presented in table~\ref{tab:results} confirm that the methodology proposed here is indeed able to capture our definition of a good set of measurements. When such set is able to distinguish the typical patterns observed in each network, its \emph{discriminability} must be high and its \emph{degree of degeneracy} is low. In this case, the set attain a large exclusion value. But being able to separate the patterns in distinct categories is not enough for a good descriptive measurement. If the measurement can truly represent any differences observed between the data elements, and it do so using the same \emph{effective resolution} for all categories, its evenness will also be high. A set of measurements achieving high exclusion and evenness values may be of great importance in understanding the building blocks of the system under study. The framework presented here can also be applied to other pattern recognition problems.

\acknowledgments
C. H. Comin thanks FAPESP (11/22639-8) for financial support. F. N. Silva acknowledges CAPES. L. da F. Costa thanks CNPq (307333/2013-2), NAP-PRP-USP and FAPESP-PRONEX (11/50761-2) for support.

\bibliography{spreadness}
\bibliographystyle{unsrt}

\end{document}